\documentclass[manuscript]{emulateapj}
\usepackage{amsmath}
\usepackage{verbatim}
\usepackage{hyperref}
\usepackage[utf8]{inputenc}
\usepackage{bm}
\usepackage{color}

\shortauthors{Evans, Laguna and Eracleous}
\shorttitle{Tidal Disruption Events With Prompt Hyperaccretion}

\def\bh#1{black hole#1
  (BH#1)\gdef\bh{BH}}
   \def\tde#1{tidal disruption event#1
  (TDE#1)\gdef\tde{TDE}}

\newcommand{\DefineRun}[2]{%
  \expandafter\newcommand\csname run-#1\endcsname{#2}%
}
\newcommand{\Run}[1]{\csname run-#1\endcsname}

\DefineRun{s1}{$\beta_{10}S_{0}M_{1}$}
\DefineRun{s2}{$\beta_{10}S_{0.65}M_{1}$}
\DefineRun{s3}{$\beta_{10}S_{-0.65}M_{1}$}
\DefineRun{s003}{$\beta_{10}S_{0}M_{0.57}$}
\DefineRun{s005}{$\beta_{10}S_{0.65}M_{0.57}$}
\DefineRun{s007}{$\beta_{10}S_{-0.65}M_{0.57}$}
\DefineRun{s004}{$\beta_{15}S_{0}M_{0.57}$}
\DefineRun{s008}{$\beta_{15}S_{0.65}M_{0.57}$}
\DefineRun{s009}{$\beta_{15}S_{-0.65}M_{0.57}$}

\begin{document}


\title{Ultra-Close Encounters of Stars With Massive Black Holes:\\ Tidal Disruption Events With Prompt Hyperaccretion}

\author{Christopher Evans\altaffilmark{1}}
\author{Pablo Laguna\altaffilmark{1}}
\author{Michael Eracleous\altaffilmark{2,1}}

\altaffiltext{1}{Center for Relativistic Astrophysics and School of
  Physics Georgia Institute of Technology, Atlanta, GA 30332}
\altaffiltext{2}{Department of Astronomy \& Astrophysics and Institute
  for Gravitation and the Cosmos, The Pennsylvania State University,
  University Park, PA 16802}

\email{plaguna@gatech.edu}

\begin{abstract} 
A bright flare from a galactic nucleus followed at late times by a $t^{-5/3}$ decay in luminosity is often considered the signature of the complete tidal disruption of a star by a massive black hole. The flare and power-law decay are produced when the stream of bound  debris returns to the black hole, self-intersects, and eventually forms an accretion disk or torus.  In the canonical scenario of a solar-type star disrupted by a $10^{6}\; M_\odot$ black hole, the time between the disruption of the star and the formation of the accretion torus could be years.  We present fully general relativistic simulations of a new class of tidal disruption events involving ultra-close encounters of solar-type stars with intermediate mass black holes. In these encounters, a thick disk forms promptly after disruption, on timescales of hours. After a brief initial flare, the accretion rate remains steady and highly super-Eddington for a few days at $\sim 10^2\,M_\odot\,{\rm yr}^{-1}$.
\end{abstract}

\maketitle

\section{Introduction} \label{sec:intro}

Observations of \tde{s} have the potential to unveil super-massive \bh{s} at the centers of galaxies. In galaxies with quiescent \bh{s}, accretion-powered nuclear activity is absent, and \tde{} signatures are, in principle, readily identifiable. The observational evidence for \tde{s} is rapidly accumulating~\citep[][]{2012Natur.485..217G,2009ApJ...698.1367G,2006ApJ...653L..25G,2011ApJ...741...73V,2014ApJ...781...59D,2011Sci...333..199L,2012ApJ...753...77C,2014ApJ...780...44C,2014ApJ...793...38A,2010ApJ...722.1035M,2013MNRAS.435.1904M}. A variety of theoretical scenarios have been considered to explain recent observations, ranging from the traditional case of a disrupted main-sequence star, to more exotic events involving the total or partial disruption of evolved stars such as a white-dwarfs~\citep[e.g.,][]{2011ApJ...726...34C,2011ApJ...743..134K}, red-giants~\citep[e.g.,][]{2005ApJ...624L..25D,2014ApJ...788...99B}, horizontal branch stars \citep[e.g.,][]{2012MNRAS.424.1268C}, and even super-Jupiters~\citep[e.g.,][]{2013A&A...552A..75N}. Of particular interest are the theoretical models predicting ignition~\citep[e.g.,][]{2009ApJ...695..404R} or the amplification of magnetic fields to launch relativistic jets \citep[e.g.,][]{2011MNRAS.416.2102G,2013ApJ...769...85S,2014ApJ...781...82C,2014MNRAS.437.2744T,2014MNRAS.445.3919K}. The motivation behind our work is that, as the number of putative \tde{} observations grows, more simulations are needed to interpret the wealth of observational data. 

\tde{} modeling was pioneered by \citet{Rees:1988bf}, \citet{1989IAUS..136..543P}, and \citet{1989ApJ...346L..13E}. These studies were first in pointing out that,  for the most likely scenario of the disruption of a main-sequence star by a $10^{6-7}\,M_\odot$  \bh{,} an UV/X-ray flare followed by a $t^{-5/3}$ decay in luminosity should be expected. The flare and decay are produced by the accretion of bound stellar debris returning to the \bh{}. In particular, the $t^{-5/3}$ decay was singled-out as a ubiquitous signature for the presence of massive \bh{s}. Subsequent studies have incorporated detailed micro-physics~\citep[][]{Lodato:2008fr,2013ApJ...767...25G,2009ApJ...695..404R}, considered a wider variety of stellar objects, such as white-dwarfs and red-giants~\citep{2013ApJ...769...85S,2012ApJ...749..117H,2004ApJ...615..855K,2009ApJ...695..404R,2014ApJ...788...99B}, covered longer dynamical times~\citep{2009ApJ...705..844G}, and included better descriptions of gravity for ultra-close encounters~\citep{1993ApJ...410L..83L,2010CQGra..27k4108R,2014PhRvD..90f4020C}. 

A challenging aspect of \tde{} studies is modeling the formation of the accretion disk~\citep{2014ApJ...784...87S,2014ApJ...781...82C,1992ApJ...385...94C,1990ApJ...351...38C}. The complication arises because, in canonical \tde{s}, the time between the disruption of the star and the formation of the accretion disk amounts to several orbital periods of the stellar debris in highly eccentric orbits. This translates into years for a solar-type star disrupted by a $10^{6}M_\odot$ \bh{.} For reference, numerical simulations in these cases cover at most a few tens of hours. 

Recent papers have addressed the circularization of the returning debris with a variety of methods.  \citet{2015arXiv150104365S} found that the debris circularizes at a larger radius than previously thought, and that the accumulation of mass in the ensuing ring is fairly slow. \citet{2015arXiv150105306G} considered the self-intersection of thin post-disruption streams for an ensemble of events and concluded that streams {\it typically} self-intersect at large distances from the \bh{,} leading to a long viscous time, hence a long delay before the onset of rapid accretion. \citet{2015arXiv150104635B} and \citet{2015arXiv150105207H} pointed out the importance of cooling on the rate of circularization of the debris, concluding that efficient cooling leads to very long circularization time scales, hence a delay in the onset of accretion. The picture emerging from the  studies is that, although accretion could be prompt under the right conditions, it is likely that the onset of accretion is delayed by an appreciable amount of time, perhaps of order a year.

Our study introduces a new class of \tde{s}, for which a puffed disk or torus forms  promptly after disruption. Furthermore, the  \bh{} accretes  at a steady and  highly super-Eddington rate of about $10^2\,M_\odot\,{\rm yr}^{-1}$ for a few days. This highly super-Eddington rate resembles those found by ~\citet{2014ApJ...781...82C}.
Our \tde{s} involve ultra-close encounters between low mass ($0.57-1\,M_\odot$) stars and a $10^5\,M_\odot$ \bh{}.  
\tde{s} with intermediate mass \bh{s}, but larger separation encounters, have been also studied by~\cite{2009ApJ...697L..77R}. Our simulations are in a regime where accounting for full general relativistic effects is needed, including those from the spin of the \bh{.} In these ultra-close encounters, the star is effectively disrupted as it arrives at periapsis, with the tidal debris plunging into the \bh{} almost instantaneously (on the timescale of one orbital period). Moreover, given the extreme proximity of the debris to the \bh{,} general relativistic precession is very efficient in circularizing stellar material to form an expanding torus. The inner material in the torus spirals into the \bh{} and is accreted at a constant rate until the supply of bound debris is exhausted. 

\section{Tidal Disruptions at a Glance} 

A star of mass $M_*$ and radius $R_*$ approaching a \bh{} of mass $M_h$, likely in a highly eccentric or parabolic orbit~\citep{Rees:1988bf,1999MNRAS.309..447M,2012EPJWC..3901004H}, will be disrupted by tidal forces 
if it wanders within a distance to the \bh{} 
\begin{equation}
\label{eq:rt}
R_t  \equiv R_* \left(\frac{M_h}{M_*}\right)^{1/3}\,,
\end{equation}
called  the \emph{tidal radius}.
It is customary to characterize the strength of a \tde{}  encounter by its \emph{penetration factor} $\beta$, which is  defined as
\begin{equation}
\label{eq:beta}
\beta \equiv \frac{R_t}{R_p}= \frac{R_*}{R_p}\left(\frac{M_h}{M_*}\right)^{1/3}\,,
\end{equation}
with $R_p$ the periapsis distance.
The fourth length scale in the problem is  the gravitational radius $R_g= G\,M_h/c^2$, which is equal to half the horizon radius for a non-spinning \bh{} and the full horizon radius for a maximally rotating \bh{}.  

\begin{figure}
\begin{center}
\plotone{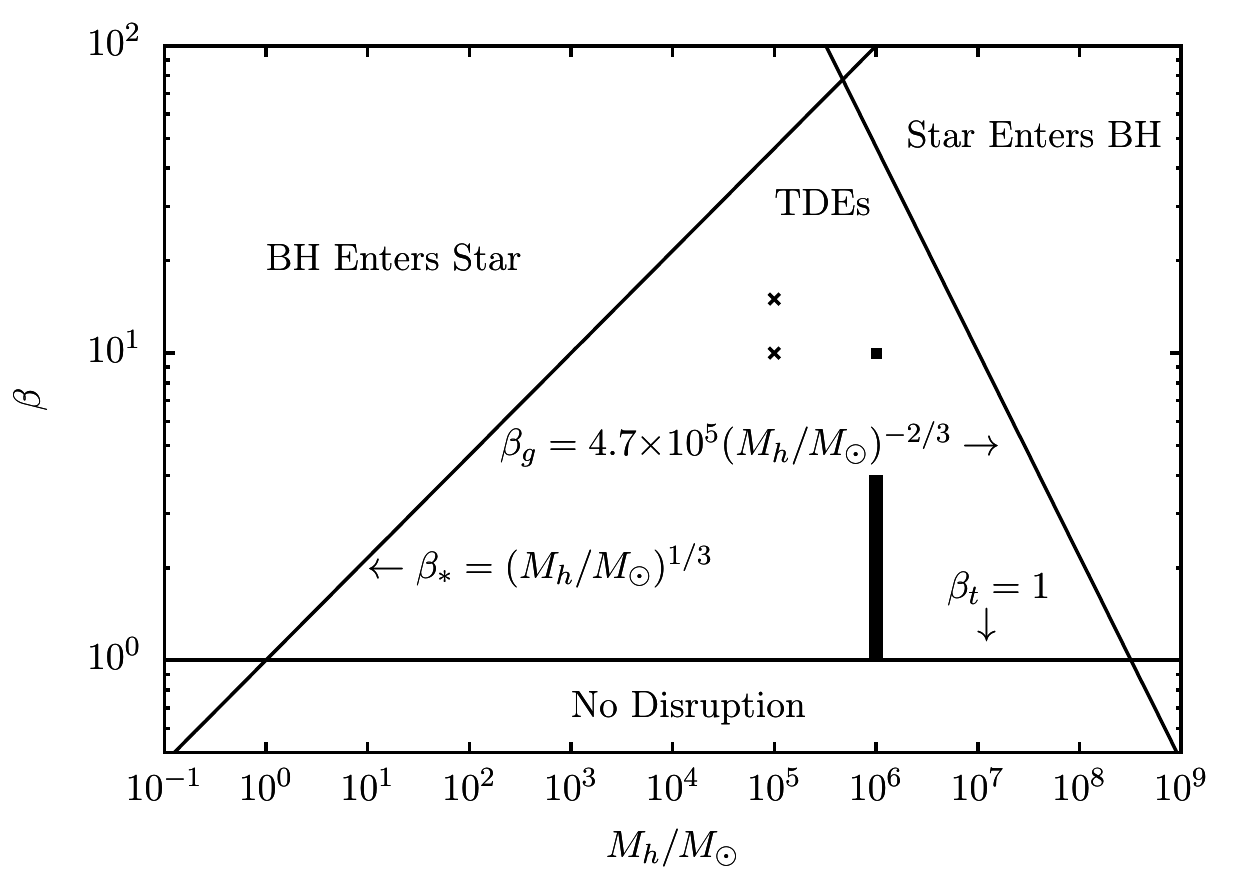}
    \caption{Tidal disruption domain for a $M_* = M_\odot$ \bh{} and $R_* = R_\odot$ star. The thick vertical line denotes canonical \tde{s}. The square point shows the ultra-close  \tde{} from~\citet{1993ApJ...410L..83L} and crosses those in the present study.}
    \label{fig:beta}
\end{center}
\end{figure}

Given $R_*,\,R_t,\,R_p$ and $R_g$, it is useful to identify the domain of astrophysical relevance of \tde{s}~\citep{1989A&A...209..103L}. In the $\beta$ {\it vs} $M_h$ plane, this domain is the triangle shown in Figure~\ref{fig:beta}, whose demarcations are obtained by interpreting $\beta$ in Eq.~(\ref{eq:beta}) as a function of $R_p$. The base of the triangle is $R_p= R_t$, i.e.  $\beta_t \equiv \beta(R_t) = 1$. Below this line ($R_p > R_t$) there is no disruption. The left side of the triangle is the line obtained by setting $R_p = R_*$; that is,
\begin{equation}
\label{eq:side1}
\beta_* \equiv  \beta(R_*) = \left(\frac{M_h}{M_*}\right)^{1/3}\,.
\end{equation}
To the left of this line ($R_p < R_*$) lie very close encounters where \emph{the \bh{} enters the star} in the process of disrupting it. This process has many similarities to the high-speed collisions of stellar-mass \bh{s} and red giants as studied by \citet{2009MNRAS.393.1016D}. Finally, the right side of the triangle is the line when $R_p = R_g$; that is,
\begin{equation}
\label{eq:side2}
\beta_g \equiv \beta(R_g) =  \frac{R_*}{G\,M_*/c^2}\left(\frac{M_h}{M_*}\right)^{-2/3} \,.
\end{equation}
To the right of this line ($R_p < R_g$) are events  where the \emph{star enters the \bh{}} before it is disrupted. 

According to Figure~\ref{fig:beta}, for a solar-type star ($M_* = M_\odot$ and $R_* = R_\odot$), the maximum \bh{} mass to disrupt the star is $M_h = 3.2 \times 10^8\,M_\odot$. Moreover, the maximum penetration factor with disruption is $\beta = 78$, which involves a $M_h = 4.7 \times 10^5\,M_\odot$ \bh{}. The edges of the triangle of astrophysical relevance are, of course, not sharp due to variations in the definition of of relevant length scales arising from the spin of the \bh{}, the  space-time curvature in the neighborhood of the \bh{,} and the internal structure of the star.

After disruption, the receding debris spreads, with roughly half of the material remaining bound to the \bh{.} The most bound material has specific binding energy given by 
\begin{eqnarray}
\label{eq:emin}
e_{\text {min}} &\simeq& -  \frac{G\,M_h\Delta R_*}{R_f^2}\nonumber\\
& \simeq& -G\,\beta^{2}f^{-2}\,\xi\,M_*^{2/3}R_*^{-1}M_h^{1/3}\,,
\end{eqnarray} 
where  $\Delta R_* = \xi\,R_*$ is the spread of the debris with $\xi$ a deformation factor, and 
\begin{equation}
R_f = \frac{R_t}{\beta} \left[ \beta\, n + (1-n)\right] = \frac{R_t}{\beta} f(\beta,n)\,
\end{equation}
is the distance to the hole when the spread in specific binding energy of the debris freezes-in. 
In the original  estimates~\citep{1989ApJ...346L..13E}, $n=0$ ($f=1$), and thus $R_f = R_p$. More recent studies~\citep{2013MNRAS.435.1809S,2013ApJ...767...25G} suggest that $n=1$ ($f=\beta$), so $R_f = R_t$.  In our study, we observed that $n \sim 0.5$, which for large $\beta$ translates to $f \sim 0.5$.

With $e_{\rm min}$ at hand, the characteristic fallback time for the most tightly bound material to return to the \bh{} is
\begin{eqnarray}
\label{eq:tmin}
t_{\text{min}} &\simeq& 2\,\pi\frac{G\,M_h}{(2| e_{\text {min}}|)^{3/2}} \nonumber\\
&\simeq& \frac{\pi}{\sqrt{2\,G}}\,\beta^{-3}\,f^{3} \xi^{-3/2}R_*^{3/2}M_*^{-1} M_h^{1/2}\,.
\end{eqnarray}
Furthermore, from the Keplerian relation 
\begin{equation}
\frac{de}{dt} = \frac{1}{3}\left(2\,\pi\,G\,M_h\right)^{2/3} t^{-5/3}
\end{equation}
 and the mass per specific binding energy 
\begin{equation}
\frac{dM_h}{de} \simeq \frac{M_*}{2\,|e_{\text{min}}|} \simeq \frac{M_*}{t_{\text{min}}^{-2/3}}\frac{1}{  \left(2\,\pi\,G\,M_h\right)^{2/3} }\,,
\end{equation}
which is assumed to be roughly constant, the accretion rate is estimated to be
\begin{equation}
\label{eq:mdot}
\dot M_h \equiv \frac{dM_h}{dt} = \frac{dM_h}{de}\frac{de}{dt} \simeq \dot M_{\max}\left(\frac{t}{t_{\text{min}}}\right)^{-5/3}\,,
\end{equation}
with $\dot M_{\rm max} \equiv M_*/(3\,t_{\rm min})$. 
The power-law decay of $t^{-5/3}$ in Eq.~(\ref{eq:mdot}) is considered to be a ubiquitous property of \tde{s}~\citep{Rees:1988bf,1989ApJ...346L..13E,1993ApJ...410L..83L,2009ApJ...695..404R}. Slight departures from this power-law have been found close to the peak accretion rate as a result of the equation of state of the star~\citep{Lodato:2008fr} or effects from the spin of the massive \bh{}~\citep{2012ApJ...749..117H}. Recent work by \citet{2013ApJ...767...25G} has shown that the canonical $t^{-5/3}$ characterizes only full tidal disruptions.  

In the case of a \tde{} with $M_* = M_\odot$, $R_* = R_\odot$ and $M_h = 10^6\,M_\odot$,
\begin{equation}
\label{eq:tmin2}
t_{\text{min}}  \simeq 0.11\,\beta^{-3}\,f^3 \xi^{-3/2}\,r_*^{3/2}\,m_*^{-1}\,M_6^{1/2}\,{\rm yr}\,,
\end{equation}
and
\begin{equation}
\label{eq:mdotmax}
\dot M_{\rm max} \simeq 0.3\,\beta^{3}\,f^{-3} \xi^{3/2} r_*^{-3/2}\,m_*^{2}\,M_6^{-1/2}\,M_\odot\,{\rm yr}^{-1}\,,
\end{equation}
where $M_6 \equiv M_h/10^6\,M_\odot$, $r_* \equiv R_*/R_\odot$ and $m_* \equiv M_*/M_\odot$. 
For comparison, the Eddington accretion rate in this situation is
$\dot M_{\text{Edd}} = 0.02\, M_6\,M_\odot\,\text{yr}^{-1}\,$
(assuming 10\% efficiency).

\section{New \tde{} Regime} 

We are interested in TDEs with $\beta =$ 10 and 15, involving a \bh{} with mass $M_h = 10^5 \,M_\odot$. They are denoted by crosses in Figure~\ref{fig:beta} and are closer to the \emph{\bh{}-enters-star} boundary than the ``canonical'' scenarios of $\beta =$ few and $M_h = 10^6 \,M_\odot$, which are denoted by a thick vertical line  in Figure~\ref{fig:beta}. In the same figure, the square point shows the ultra-close \tde{} from~\citet{1993ApJ...410L..83L}. Here, we consider non-spinning \bh{s} and \bh{s} with spin $a/M_h = \pm 0.65$. The sign denotes whether the spin of the \bh{} is aligned (plus) or anti-aligned (minus) with the orbital angular momentum of the star. The disruptions involve main-sequence stars injected in parabolic orbits with masses $M_* = 1\,M_\odot$ and $0.57\,M_\odot$, modeled as polytropes with equation of state $p \propto \rho_0^\Gamma$ and $\Gamma = 4/3$. During the evolution, we use a gamma-law equation of state $p = \rho_0\epsilon(\Gamma-1)$. Table~\ref{table:parms} provides the parameters of the simulations: the penetration factor $\beta$, the \bh{} spin parameter  $a$, and the mass of the star $M_*$.  

The simulations were carried out with our Maya code and fully account for general relativistic effects. This code was also used in our previous general relativistic \tde{} study~\citep{2012ApJ...749..117H}. The Maya code is 4th order accurate in solving the Einstein equations and 2nd order in the hydrodynamics equations. The general convergence of the code is slightly above 2nd order. We employ nine levels of mesh refinements. All but the coarsest mesh have $70^3$ grid-points, with the coarsest having $120^3$. The resolution on the finest mesh is $R_g/50$.

\begin{table*}[htbp]
\begin{ruledtabular}
    \caption{Simulation Parameters and Accretion Rates.}
    \label{table:parms}
    \begin{tabular}{lccccc}
    Run & $\beta$ & $a/M_h$ & $M_*/M_\odot$& $\dot M_{\rm max}$ ($M_{\odot}\,{\rm yr}^{-1}$) & $\dot M_{\rm late}$($M_{\odot}\,{\rm yr}^{-1}$)  \\
\hline
  \Run{s1}   & 10& 0 & 1& $3.6 \times 10^{2}$ & $1.0 \times 10^{3}$  \\ 
   \Run{s2}  &10 &0.65 & 1 & $1.0 \times 10^{4}$ & $7.5 \times 10^{2}$ \\ 
   \Run{s3} & 10& -0.65 &1& $1.2 \times 10^{5}$ & $2.0 \times 10^{2}$ \\
    \Run{s003}  & 10 & 0 & 0.57 & $3.5 \times 10^{2}$ & $1.0 \times 10^{2}$  \\ 
  \Run{s005}    & 10 &0.65& 0.57& $7.6 \times 10^{3}$ & $4.0 \times 10^{2}$  \\ 
   \Run{s007}   & 10& -0.65 & 0.57& $7.9 \times 10^{4}$ & $3.0 \times 10^{2}$  \\ 
 \Run{s004}   & 15 & 0 & 0.57 & $2.0 \times 10^{4}$ & $4.0 \times 10^{2}$\\ 
   \Run{s008}  & 15 & 0.65 & 0.57 & $8.8 \times 10^{4}$ & $3.5 \times 10^{2}$\\
   \Run{s009}   & 15 & -0.65 & 0.57 & $3.8 \times 10^{5}$ & $2.0 \times 10^{2}$\\ 
    \end{tabular}
\end{ruledtabular}
\end{table*}

In general terms, the \tde{s} studied here involve stars comparable in size to the \bh{}, $R_* \simeq 4.7\,M_5^{-1}\,r_*\,R_g$, periapsis distances of about  
\begin{equation}
R_p \simeq 4.64 \,\beta_{10}^{-1}\,M_5^{1/3}\,m_*^{-1/3}\,R_*,
\end{equation}
and tidal radius 
\begin{equation}
R_t\simeq 218\,M_5^{-2/3}\,m_*^{-1/3}\,r_*\,R_g\,,
\label{eq:rt2}
\end{equation}
where $M_5 \equiv M_h/10^5\,M_\odot$ and $\beta_{10} \equiv \beta/10$. 
Although the value of $R_p$ suggests that the star could potentially swing by the \bh{} without the \bh{} entering the star, the combination of large $\beta$ and general relativistic effects produce an outcome dramatically  different from the situations with $\beta \sim 1$ and $10^6\,M_\odot$ mass \bh{s} (next section).

With $\beta \ge 10$ and intermediate mass \bh{s}, the star will be effectively disrupted and stretched to a few times its original size by the time it reaches periapsis passage. A deformation factor $\xi \simeq 4$ was found to be common in our simulations. Therefore, from Eqs.~(\ref{eq:tmin}) and (\ref{eq:mdot})
\begin{equation}
\label{eq:tmin3}
t_{\text{min}}  \simeq 137\,\beta_{10}^{-3}\,f^{3} \xi_4^{-3/2}\,r_*^{3/2}\,m_*^{-1}\,M_5^{1/2} \,{\rm s}\,,
\end{equation}
and
\begin{equation}
\label{eq:mdotmax2}
\dot M_{\rm max} \simeq  9.6\times 10^3\,\beta_{10}^{3}\,f^{-3} \,\xi_4^{3/2}\,r_*^{-3/2}\,m_*^2\,M_5^{-1/2}\,M_\odot\,{\rm yr}^{-1}\,,
\end{equation}
where $\xi_4 = \xi/4$. For reference, the Eddington accretion rate in this case is $\dot M_{\text{Edd}} = 0.002\, M_5\,M_\odot\,\text{yr}^{-1}$. Our simulations  show that $\dot M_{\rm max} \sim 10^4 \,M_\odot\,{\rm yr}^{-1}$ and $t_{\rm min} \sim 25\,{\rm s}$. This time-scale is  comparable to the circular orbital period of the most bound material: 

\begin{eqnarray}
P_{\rm circ}&=& 2\,\pi\sqrt{\frac{(R_p-\Delta R_*)^3}{G\,M_h}} \nonumber\\
& \simeq& 316 \,(1-\Delta R_*/R_p)^{3/2}\,\beta_{10}^{-3/2}\,r_*^{3/2}\,m_*^{-1/2}\,{\rm s} \nonumber\\
&\simeq&16.5\,\beta_{10}^{-3/2}\,r_*^{3/2}\,m_*^{-1/2}\,{\rm s}\,,
\end{eqnarray} 
where $\Delta R_*/R_p \simeq 0.86\,\xi_4\,\beta_{10}\, m_*^{1/3}\,M_5^{1/3}$.

\section{Anatomy of a Disruption}  

\begin{figure}
\begin{center}
\includegraphics[scale=0.65]{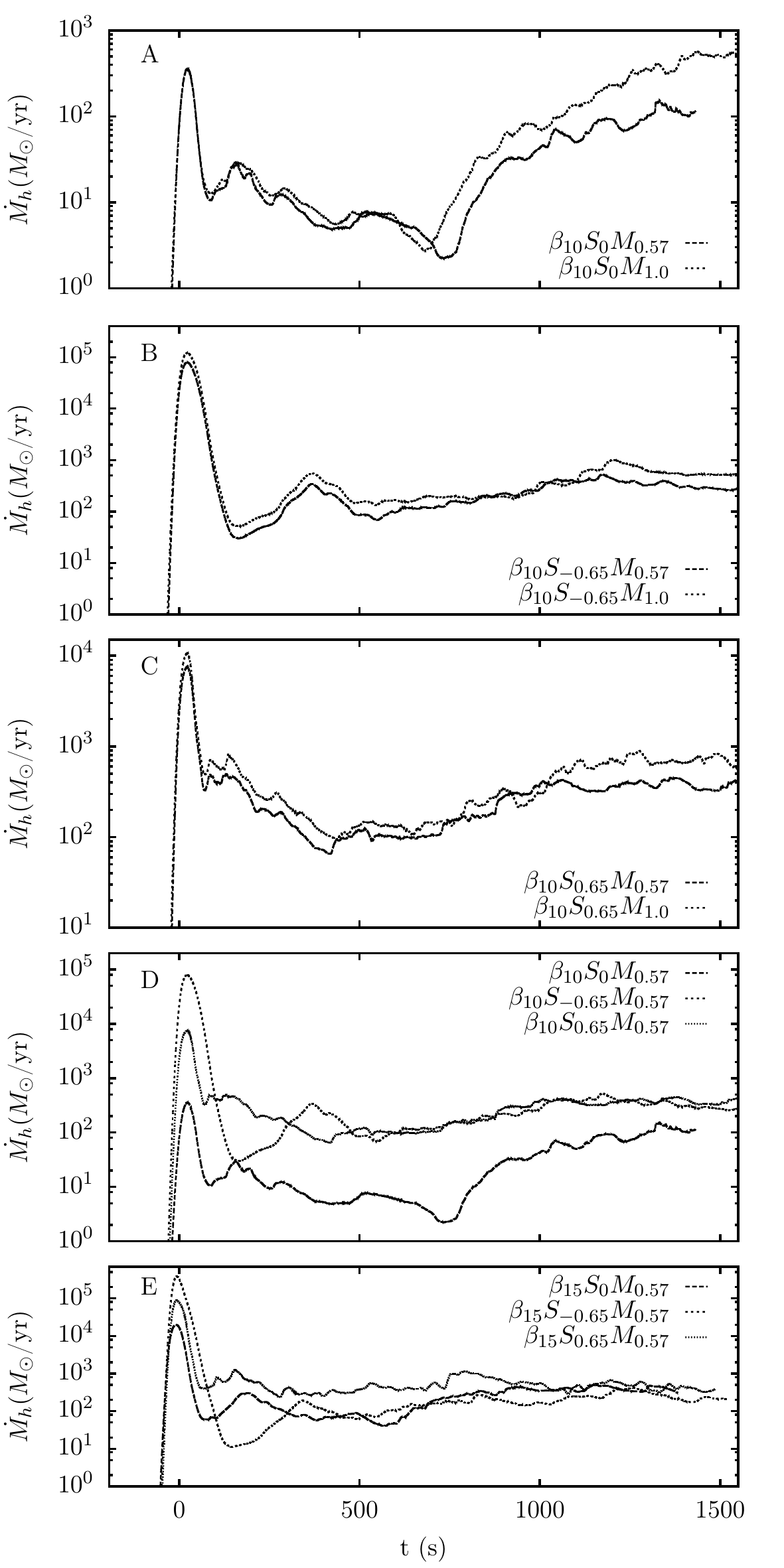}
    \caption{\bh{} accretion rate as a function of time. Panels A-C display cases with penetration factor $\beta=10$ grouped by the spin parameter of the \bh{}, with $a/M_h = 0, -0.65$ and 0.65 corresponding to panels A, B and C, respectively. The remaining panels display accretion rates for $M_* = 0.57\,M_\odot$ with penetration factor $\beta=10$ (panel D) and $\beta = 15$  (panel E).}
    \label{fig:mass-effect}
\end{center}
\end{figure}

We focus this discussion chiefly on the accretion rates onto the \bh{.} Figure~\ref{fig:mass-effect} shows the accretion rates of tidal debris through the \bh{} horizon. The time axis is such that $t=0$ s denotes periapsis passage. Three distinct accretion epochs or stages are identifiable in most of the cases. 

The first phase is a narrow spike or flare in the accretion rate. The spike is due to the portion of the stellar debris that immediately plunges into the \bh{.} An estimate of this accretion rate is given by
\begin{eqnarray}
\label{eq:bondi}
\dot M_h &\sim& A_h\,\rho_\infty\,v_\infty\nonumber\\
& \sim& 10^4\,\beta_{10}^{1/2}\,M_5^{7/3}\,m_*^{7/6}\,r_*^{-7/2}\,M_\odot\,\text{yr}^{-1}\,,
\end{eqnarray}
where we have used $A_h \sim 4\,\pi\,R^2_g$, $v_\infty \sim (G\,M_h/R_p)^{1/2}$ and $\rho_\infty \sim M_*/(4\,\pi\,R_*^3/3)$. This estimate is consistent 
with the values for $\dot M_{\rm max}$ reported in Table~\ref{table:parms}.

After the accretion flare there is a decay phase, which for a non-spinning \bh{} (see top panel in Figure~\ref{fig:mass-effect}) loosely resembles a power-law decay. The duration of this decay seems to depend on the spin of the  \bh{} and the penetration factor $\beta$. 

Finally, as new material returns to the \bh{,} it circularizes and forms an accretion torus  on a timescale of $\sim 10\,t_{\rm min} \sim 1,400\,{\rm s}$~\citep{1989ApJ...346L..13E}, with the accretion eventually reaching steady state at about $\dot M_{\rm late} \sim 10^2 M_\odot\,\text{yr}^{-1}$, as noted in the last column of Table~\ref{table:parms}. The steady state accretion will cease when the mass supply from the bound debris is depleted. 

\begin{figure*}[htbp]
   \centering\plotone{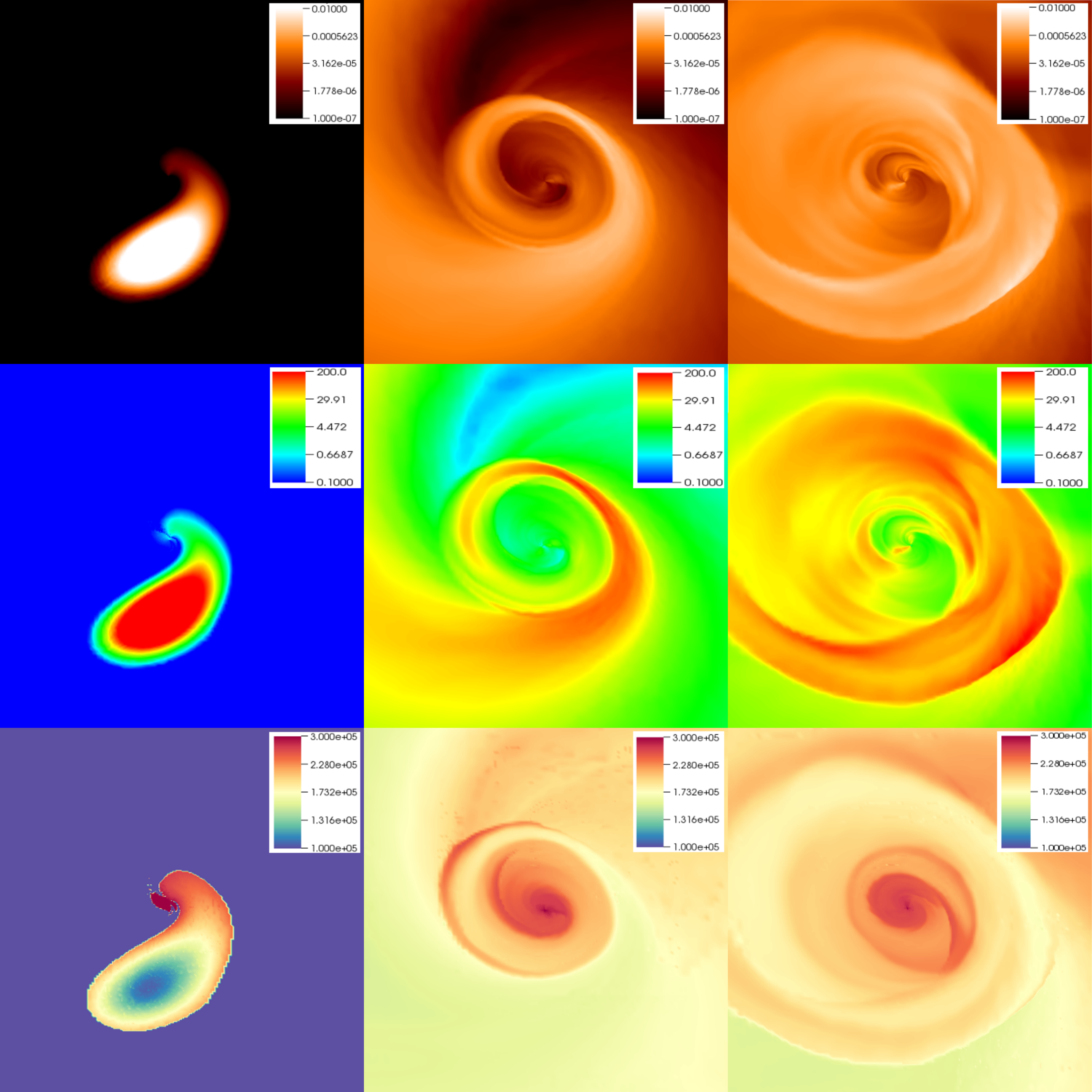}
    \caption{Snapshots of the density (top row), temperature (middle row) and specific entropy (bottom row) of the stellar debris for the \Run{s003} case. Columns from left to right depict times $t = -28$ s, 450 s and 1,250 s, respectively.}
    \label{fig:composite_10}
\end{figure*}

\begin{figure}
\begin{center}
\includegraphics[scale=0.4]{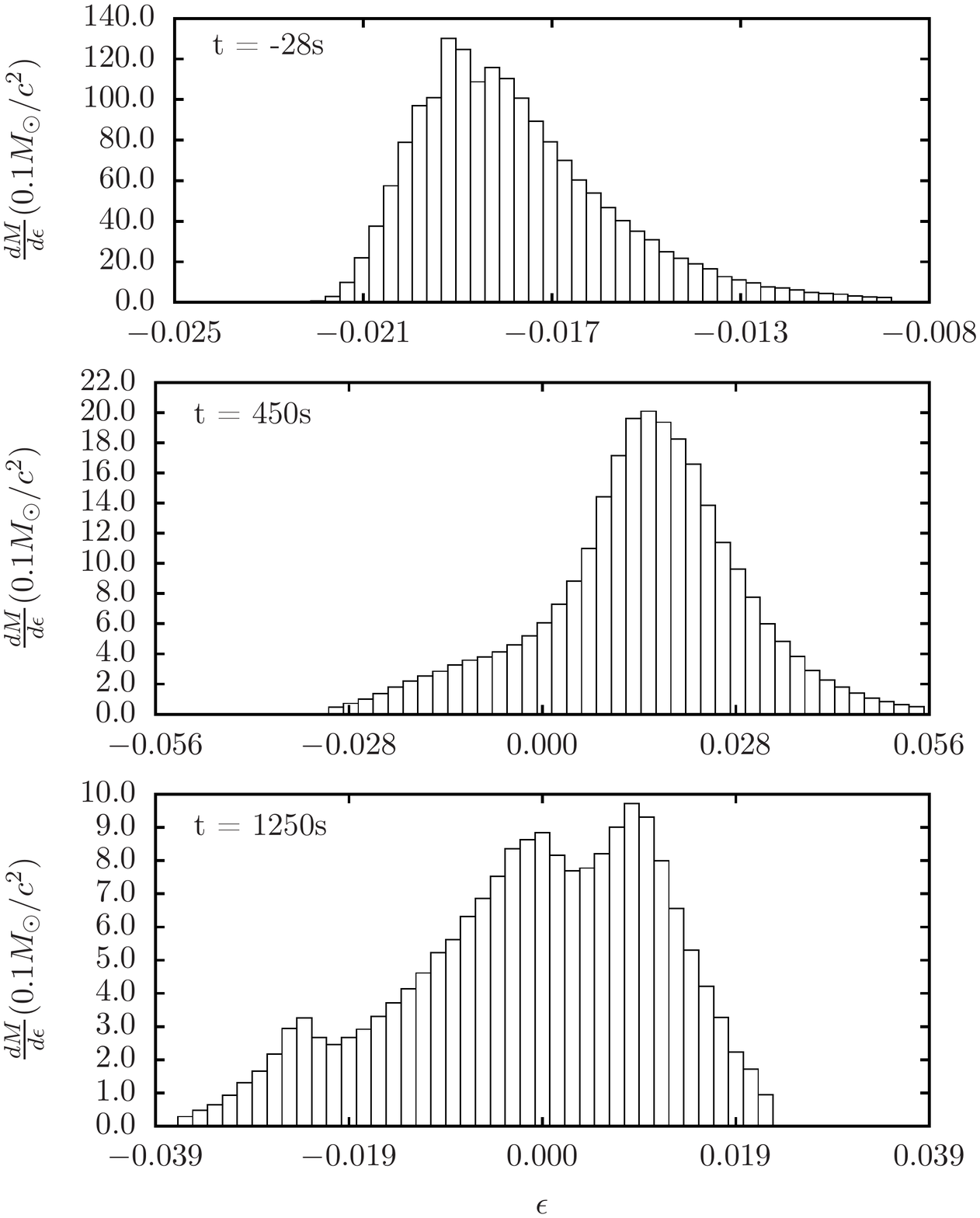}
    \caption{Histograms of the mass per unit binding energy as a function of binding energy for the case in Figure~\ref{fig:composite_10}.}
    \label{fig:massdistribution}
\end{center}
\end{figure}

\section{Simulation Results} 

To illustrate the effect of the mass of the star, panels A, B and C in Figure~\ref{fig:mass-effect}  show the \bh{} accretion rate for $\beta =10$, grouping the cases with the same \bh{} spin parameter. Since the accretion rate seems to be insensitive to the mass of the star, we will  focus on the $M_* = 0.57\,M_\odot$ star given that we have a wider variety of simulations for this star.

In Figure~\ref{fig:mass-effect}, we also organize the runs according to $\beta$, wit $\beta = 10$  in panel D, and $\beta=15$ in panel E. The peak accretion rate of the flare is higher if the \bh{} is spinning. Interestingly, the case with a counter-rotating orbit (i.e. \bh{} spin anti-aligned with the orbital angular momentum) yields the largest peak accretion rate. For the $\beta = 10$ simulations, the post-flare accretion rate depends on the \bh{} spin magnitude but not its orientation, which is consistent with similar findings for the steady-state, subsonic accretion onto a moving \bh{}~\citep{PhysRevLett.60.1781}. The late-time accretion rate seems to increase with the spin of the \bh{.}   Another difference between the $a/M_h = 0$ and the $a/M_h = \pm 0.65$ cases is that for the latter, the time scale for flare decay is shorter; that is, the late constant accretion phase, which signals the formation of the torus, is reached after a couple of hundred seconds. 

The corresponding $\beta=15$ cases show that the spin of the \bh{} does not play a role in determining the late-time, constant accretion rate. Furthermore, as with the $\beta=10$ cases with spinning \bh{s}, the constant accretion phase is reached within a few hundred seconds. For both the $\beta = 10$ and $\beta=15$ cases, the late-time accretion rate is $\sim 10^2\,M_\odot\,{\rm yr}^{-1}$.

In summary, we found that the accretion rate is not very sensitive to $M_*$, and that the rate reached during the flare depends on both  $a/M_h$ and $\beta$. Furthermore, the late accretion rate for $\beta =15$ is not sensitive to the spin of the \bh{}, and for both penetration factors is approximately $\dot M\sim 10^2\,M_\odot\,{\rm yr}^{-1}$. With the exception of the cases with $\beta=10$ and $a/M_h = 0,\,0.65$ (panels A and C in Figure~\ref{fig:mass-effect}), there are no hints of a power-law decay rate.   

Figure~\ref{fig:composite_10} shows snapshots of the density (top row), temperature (middle row) and specific entropy (bottom row) of the stellar debris for the \Run{s003} case. Columns from left to right depict times $t = -28$ s, 450 s and 1,250 s, respectively.  Self-intersection of debris and subsequent formation of the accretion torus are evident. The increase in temperature and entropy as a result of shocks due to self-intersection of material is also evident.

Finally, Figure~\ref{fig:massdistribution} shows histograms of the mass per unit binding energy as a function of binding energy for the same case and times as in Figure~\ref{fig:composite_10}. It is evident that $dM_h/d\epsilon$ is not flat, as is required to achieve the $t^{-5/3}$ power-law decay.

\citet{1982Natur.296..211C} were first to suggest a tidal compression perpendicular to the orbital plane $\propto$  $\beta^{-3}$ that would yield an increase in the central density and temperature of the star $\propto$ $\beta^3$ and $\beta^2$, respectively. We do not observe such extreme compression. As mentioned before, the star arrives at periapsis effectively disrupted, with its central density decaying and undergoing only a mild compression, adequately handled with the resolution used in our simulations.  

\section{Conclusions}  

We presented a new class of \tde{s} showing prompt formation of an accretion torus and hyperaccretion.
These \tde{s} involve ultra-close encounters with a $M_h =10^5\,M_\odot$ \bh{}. The accretion rates are highly super-Eddington. Additionally, there is no evidence of a $t^{-5/3}$ decay in the accretion rate. This is likely due the strong influence of general relativistic effects in these ultra-close encounters. The late-time accretion rate, once the torus has formed, reaches an approximate steady-state and remains highly super-Eddington at $\dot M_{\rm late} \sim 10^2 M_\odot\,\text{yr}^{-1}$. With this rate, the \bh{} should be able to accrete the majority of the bound tidal debris in just a few days. A word of caution is needed regarding the accretion rate quotes in this study. Our simulations did not include effects from radiation, which could potentially decrease the rates. However, it is not likely that radiation effects would diminish the rates enough to make them sub-Eddington, since the Eddington rate in these situations is  $\dot M_{\text{Edd}} = 0.002\, M_5\,M_\odot\,\text{yr}^{-1}\,$ (assuming 10\% efficiency).

In a subsequent study, we will expand the parameter space of our simulations and investigate the emission properties of the tidal debris. It will be important to investigate further the role of shocks in heating and circularizing the debris as well as the overall role of cooling effects. Moreover, we will consider inclined orbits relative to the spin axis of the \bh{} in order to compare our results with those of ~\citet{2014ApJ...784...87S}. We are also interested in the possibility of amplification of magnetic fields and whether the class of \tde{s} in the present study could provide an explanation for jetted events, such as Swift~J1644+57~\citep{2011Natur.476..421B,2011Sci...333..199L}.

We thank Tamara Bogdanovi\'c. Work supported by NSF 1205864, 1212433, 1333360. Computations at XSEDE TG-PHY120016.

\bibliographystyle{astroads}

\end{document}